\input{aipcheck}

\newcommand{\numu}{\nu_\mu}
\newcommand{\numub}{\bar{\nu}_\mu}

\newcommand{\mup}{\mu^{+}}
\newcommand{\mum}{\mu^{-}}

\newcommand{\pip}{\pi^{+}}
\newcommand{\pim}{\pi^{-}}
\newcommand{\thetmu}{\theta_{\mu}}

\newcommand{\uz}{\,cos\, \theta_\mu}

\newcommand{\tmu}{T_\mu}

\newcommand{\dbldiffl}{\frac{d^2\sigma}{dT_\mu d\uz}}

\documentclass[
    ,final            
  ]
  {aipproc}

\layoutstyle{8x11single}
\usepackage{multirow}

\begin{document}

\title{New Anti-Neutrino Cross-Section Results from MiniBooNE}


\author{Joseph Grange$^*$ and Ranjan Dharmapalan$^\dag$ for the MiniBooNE Collaboration}{
  address={\centering $^*$University of Florida, Gainesville FL \\ \centering $^\dag$University of Alabama, Tuscaloosa AL \centering}
}



\begin{abstract}

The first measurements of antineutrino charged-current quasielastic ($\numub$ CCQE, $\numu + N \to \mup + N'$) and neutral-current elastic ($\numub$ NCE, $\numu + N \to \numu + N$) cross sections with  $\langle E_{\bar{\nu}} \rangle$ $<$ 1 GeV are presented.  To maximize the precision of these measurements, many data-driven background measurements were executed, including a first demonstration of charge separation using a non-magnetized detector.  Apart from extending our knowledge of antineutrino interactions by probing a new energy range, these measurements constrain signal and background processes for current and future neutrino oscillation experiments and also carry implications for intra-nuclear interactions.

\end{abstract}

\maketitle


\section{Introduction}


The Mini Booster Neutrino Experiment (MiniBooNE) has collected over 1.7 $\times$ 10$^{21}$ protons-on-target (POT) across the neutrino ($\langle E_{\nu} \rangle$ = 788 MeV) and antineutrino-mode ($\langle E_{\bar{\nu}} \rangle$ = 665 MeV) run configurations.  Cross sections for channels contributing roughly 90\% of the total event rate for neutrino-mode data have been published~\cite{qePRD,CCpip,NCpi0,CCpi0,nuNCE}.  Observed enhancements in some channels relative to predictions from the Relativistic Fermi Gas (RFG) model~\cite{RFG} have led to speculations that nuclear effects may be greater than previously expected~\cite{qeReview}.

A strong test of the underlying physics of the interactions contributing to the MiniBooNE data sets is available with comparisons in the behavior between neutrinos and antineutrinos.  In particular, combined measurements such as cross-section ratios contain information from both processes and also allow more precise measurements through the exploitation of correlated systematic uncertainties.  At this conference, first results of cross-section measurements of $\numub$ CCQE, NCE, and various combined quantities of these cross sections with the corresponding $\numu$ processes were presented and are described here.  These results provide important constraints on signal and background processes for current and future neutrino oscillation experiments through the substantial overlap in the observed neutrino and antineutrino energy spectra.  These experiments include NO$\nu$A~\cite{nova}, T2K~\cite{t2k}, and LBNE~\cite{lbne}.

\section{The $\numu$ Background}

The motivation for and the execution of the measurements presented in this section are described in detail elsewhere~\cite{wsPRD,nubDblDiffl}.

\begin{figure}
$
\begin{array}{cc}
\includegraphics[scale=0.35]{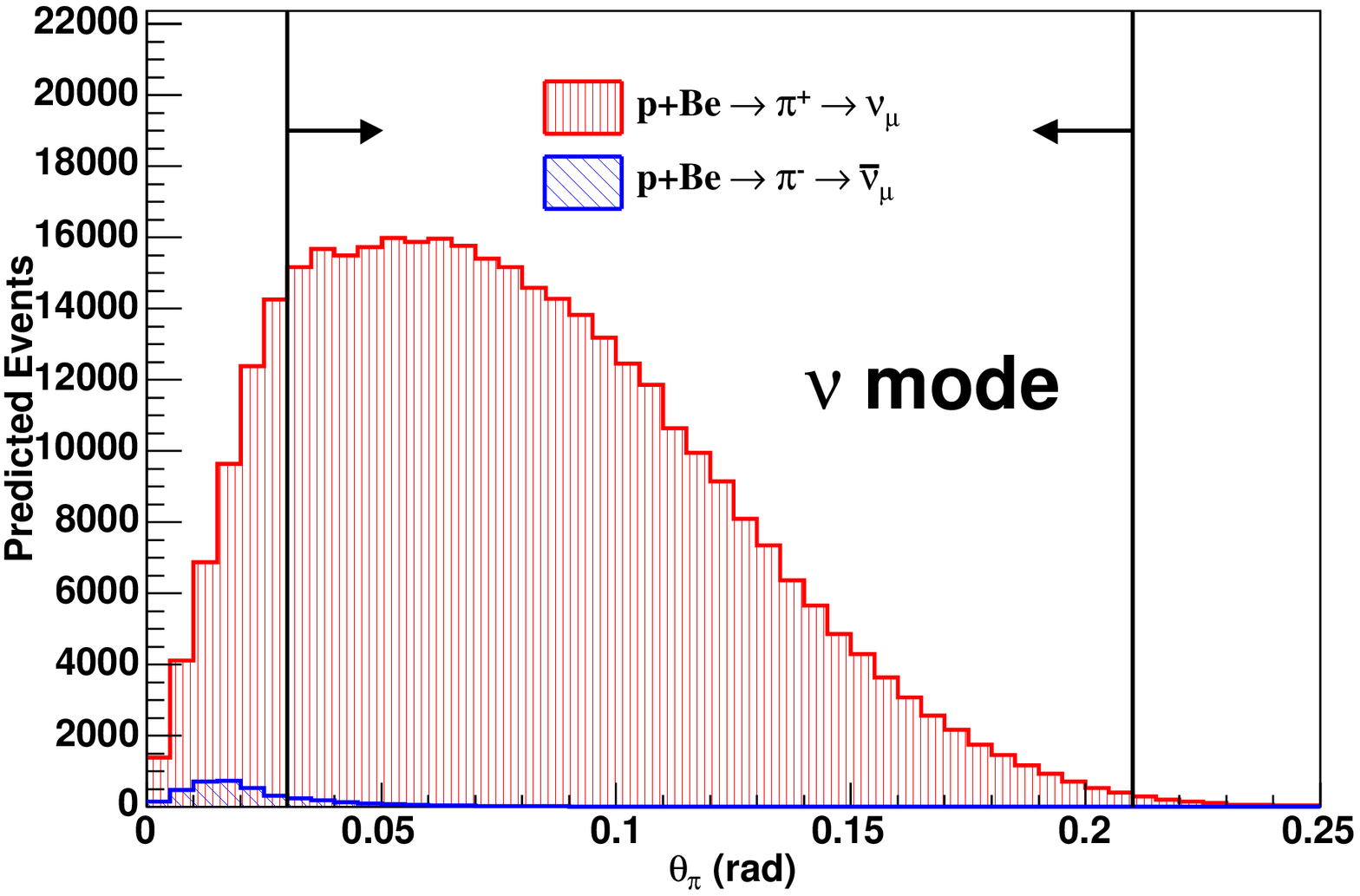} &
\includegraphics[scale=0.35]{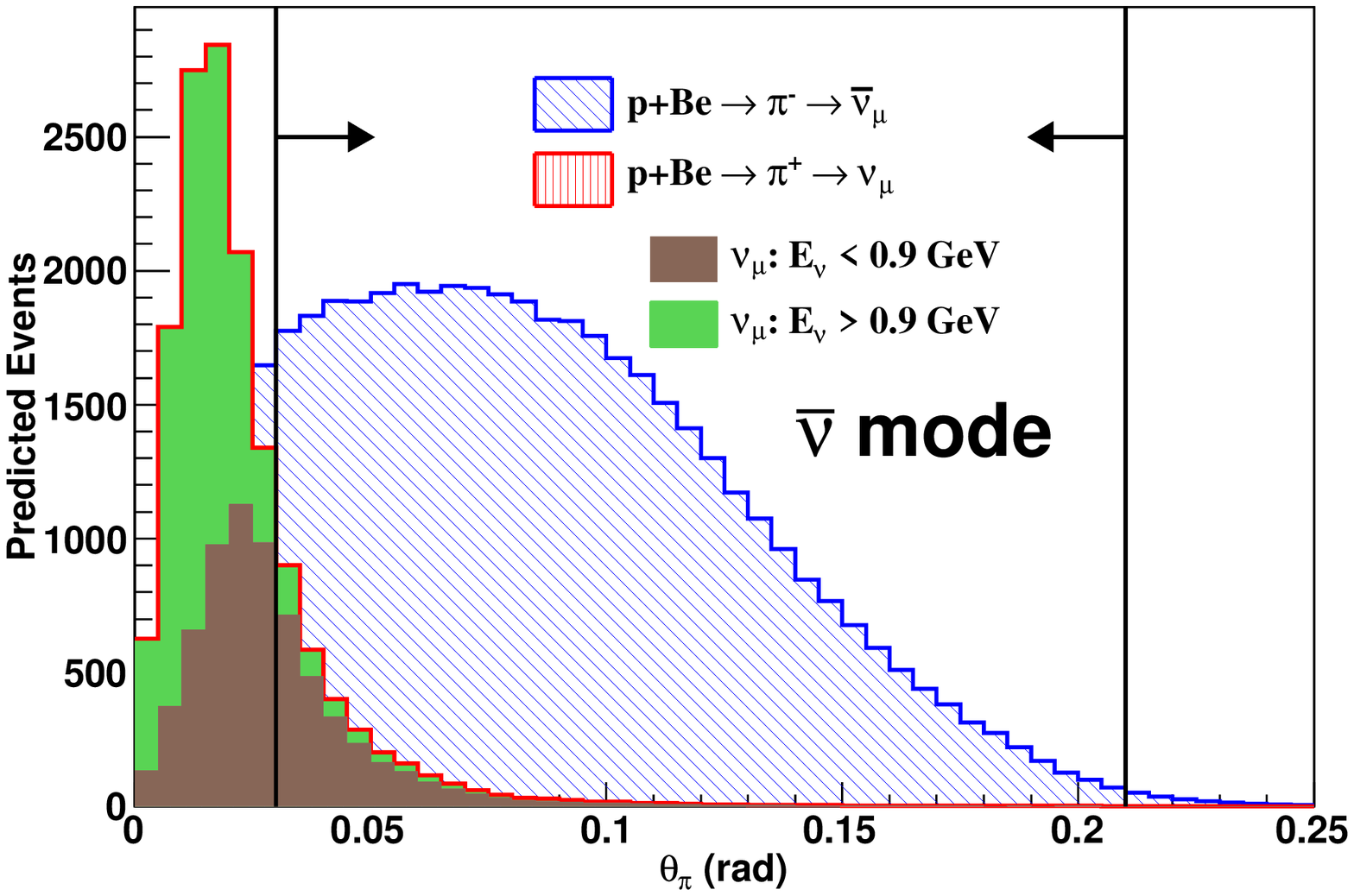} \\
\end{array}$
\caption{\label{fig:thPi} Predicted angular distributions of the pion scattering angle ($\theta_{\pi}$) producing $\numu$ and $\numub$ in neutrino (left) and antineutrino (right) modes.  Only pions leading to $\numu$ and $\numub$ events in the detector are shown, and all distributions are normalized to 10.1 $\times \,\, 10^{20}$ protons on target.  Arrows indicate the regions where HARP data are available.  Figures adapted from Ref.~\cite{wsPRD}.}
\end{figure}

Precision $\numu$ and $\numub$ cross sections with the MiniBooNE detector would not be possible without dedicated hadroproduction data from the HARP~\cite{HARP} experiment.  However, these data do not constrain small regions of $\pi$ kinematics important to the prediction of $\numu$ in the anti-neutrino mode beam.  Figure~\ref{fig:thPi} shows the production angle of $\pip$ and $\pim$ relative to the incoming proton beam at the MiniBooNE target contributing to the neutrino and antineutrino-mode data sets.

The particulars of Figure~\ref{fig:thPi} warrant a few remarks:

\begin{itemize}


\item both parent pion distributions leading to the ``wrong-sign" contribution (neutrinos in antineutrino mode and {\it vice versa}) peak at the lowest opening angles.  This shows how these events contribute to the beam: wrong sign pions at small angles are not deflected by the horn's magnetic field.

\item the antineutrino contribution to the neutrino-mode data is negligible in comparison to the converse.  This is due to a convolution of flux and cross-section effects that simultaneously serve to enhance the neutrino component and suppress the antineutrino contribution: the leading-particle effect at the beryllium target (the $p + \textrm{Be}$ initial state has a net positive charge) naturally leads to the creation of roughly twice as many $\pip$ as $\pim$, and neutrino cross sections are typically around three times as large as antineutrino cross sections around 1~GeV.
\item the above observation explains why this is a complication unique to antineutrino mode: the wrong-sign component in neutrino-mode data is small enough so that even for large fractional uncertainty on this background, the resultant error on the $\numu$ cross-section measurements are negligible compared to other systematic uncertainties.  
\item as seen in the antineutrino-mode distribution, high-energy $\numu$'s are strongly correlated with the decay of $\pip$ created at very small opening angles.  This indicates their flux is more poorly constrained by the HARP data compared to lower-energy $\numu$'s.  So, not only is the overall $\numu$ flux in antineutrino mode largely unconstrained, the accuracy of the $\numu$ flux prediction may be a function of neutrino energy.

\end{itemize}

This motivates dedicated studies of the $\numu$ contribution to the antineutrino-mode beam.  The MiniBooNE detector is not magnetized, so the contributions from $\numu$ and $\numub$ cannot be separated in charged-current samples based on the observed charge of the outgoing lepton.  Therefore, statistical asymmetries in the way neutrinos, antineutrinos and their byproducts interact in the detector are exploited to directly constrain the contribution from neutrinos.  Three analyses, based on independent asymmetries, are executed:

\begin{enumerate}

\item $\numu$ CC$\pip$ ($\numu + N \to \mum + N + \pip$) events typically yield a $\mum$ and two electrons, one from the $\mum$ decay and another from the $\pip \to \mup \to \,$e$^+$ chain.  A second electron in the $\numub$ CC$\pim$ ($\numub + N \to \mup + N + \pim$) process is typically not created, due to $\sim$ 100\% nuclear capture of $\pim$ on carbon~\cite{pimCap}.  Therefore, a simple requirement of the presence of a single muon and two electrons gives sensitivity to the $\numu$ content of the beam.

\item Due to the $\sim$ 8\% rate of nuclear capture for $\mum$ in the presence of carbon~\cite{pimCap}, $\numu$ charged-current interactions are less likely to yield a decay electron compared to the $\numub$ processes.  Inclusive charged-current samples consisting of a single muon and one or two electrons are simultaneously adjusted to give consistency with the observed samples.

\item The kinematics of the $\mup$ created in $\numub$ CCQE interactions are predicted to be much more forward-peaked relative to the incoming neutrino direction, compared to the $\mum$ in the $\numu$ CCQE process.  The muon angular distribution was fit to a linear combination of the $\numu$ and $\numub$ contributions.

\end{enumerate}

Note the third analysis is dependent on knowing the kinematics of the $\mup$ in $\numub$  interactions contributing to the MiniBooNE CCQE sample.  As measurements of this quantity are a primary goal of the $\numub$ cross-section measurements, the results of the third technique are {\it not} used to subtract the $\numu$ background in determining the $\numub$ CCQE and NCE cross sections.  However, this technique could prove to be powerful once the kinematics of these interactions are better understood.

\begin{figure}
  \includegraphics[scale=0.45]{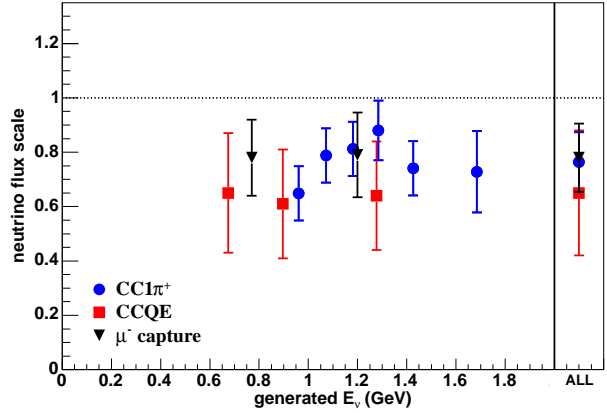}
  \caption{Summary of the results from three techniques used to measure the $\numu$ flux in the antineutrino-mode beam.  Measurements performed in exclusive regions of reconstructed energy are placed here at the mean of their associated distribution of true energy.  Shown as a dotted line at unity, the measurements are made relative to an extrapolation of HARP data into a region where no relevant hadroproduction data exists.}
  \label{fig:wsSumm}
\end{figure}

The MiniBooNE-measured $\numu$ cross sections dominant in these interaction samples are applied to the simulation, and so the rate analyses executed based on the above asymmetries also test the accuracy of the unconstrained $\numu$ flux prediction.  To gain sensitivity to the accuracy of the flux spectrum, these analyses are binned as finely as possible, as allowed by the collected statistics, in regions of reconstructed neutrino energy.  Figure~\ref{fig:wsSumm} summarizes the results from the three analyses of the $\numu$ flux in the antineutrino-mode beam.  The data prefers a uniform reduction of $\sim$ 20\% relative to the prediction that has been extrapolated into a kinematic region not directly constrained by HARP data.


These analyses are a first demonstration of charge separation in the absence of a magnetic field.  The techniques could be used in present nonmagnetized neutrino detectors, and could inform design choices in future experiments.

With the largest background to the antineutrino-mode data directly constrained by complementary {\it in situ} measurements, we can turn our attention to the analysis of antineutrino interactions.

\section{$\numub$ CCQE double-differential cross section}

The $\numub$ CCQE cross-section analysis and results are described in greater detail in Ref.~\cite{nubDblDiffl}.

The $\numub$ CCQE selection is based exclusively on simple kinematic requirements of the prompt muon and the presence of its decay positron.  Signal events are predicted to account for $\sim$ 60\% of the analysis sample, where $\numu$ events account for about 20\% of the selected events, and the contribution from CC$\pim$ is $\sim$ 15\%.


The largest background of $\numu$ interactions is constrained by the measurements described in the previous section, and the second-largest background of $\numub$ CC$\pim$ interactions also deserves to be addressed.  Due to $\sim$ 100\% nuclear $\pim$ capture, these events these events do not produce a second electron and hence are not separable from the $\numub$ CCQE sample.  In the absence of the ability to directly constrain these interactions, the constraint of $\numu$ CC$\pip$ interactions measured in the MiniBooNE neutrino-mode data~\cite{wsPRD} is extrapolated and applied to the $\numub$ CC$\pim$ processes.  Consistency between this prediction and a modern model for single-pion production~\cite{jarekProc} that successfully describes the bulk of world pion-production data suggests the assigned uncertainty of $\sim$ 20\% is sufficient.

\begin{figure}
\includegraphics[trim=0cm 0cm 3cm 0cm,clip=true,scale=0.90]{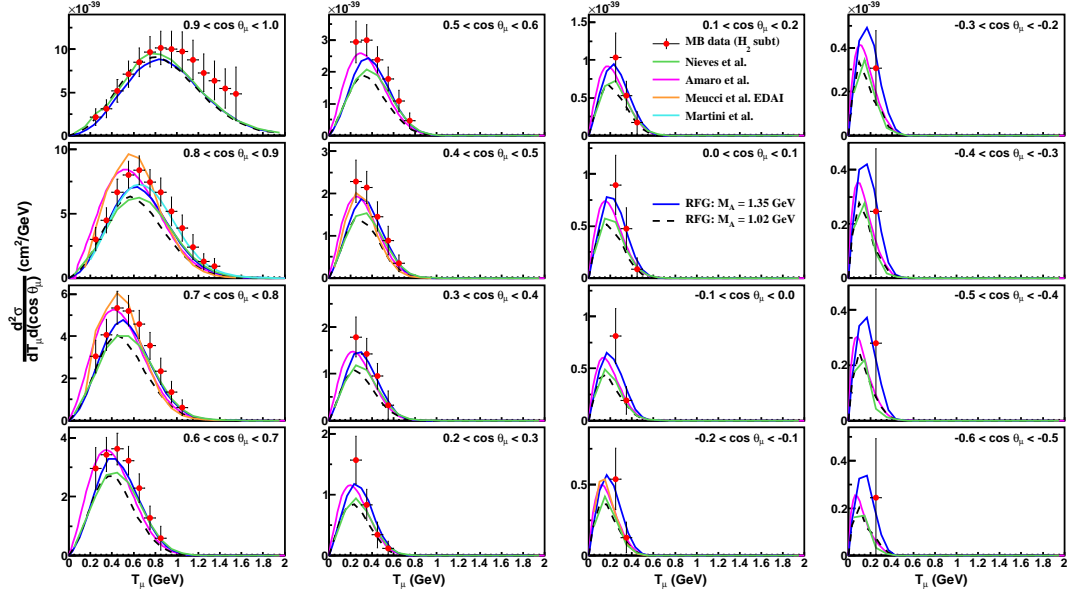} 
\caption{\label{fig:dblDiffl} Projections of the double-differential cross section for $\numub$ CCQE on carbon in muon kinetic energy $\tmu$ for various scattering angles $\uz$.  The Martini {\it et al.} work~\cite{martini} also shows projections of the double-differential cross section as a function of scattering angle.}
\end{figure}

With the most important backgrounds directly or indirectly constrained by {\it in situ} measurements, they may be reliably subtracted from the data to calculate the $\numub$ CCQE cross sections.  The main result of this work is the minimally-model dependent double-differential cross section $\dbldiffl$, where $\tmu$ ($\thetmu$) is the muon kinetic energy (scattering angle).  The flux-integrated double-differential cross section per nucleon in the $i^{\textrm{th}}$ kinematic region is given by:

\begin{center}
\begin{equation}
\label{eqn:dblDiffl}
\left(\frac{d^2\sigma}{d\tmu\,d\left[\uz\right]}\right)_i = \frac{\sum_j U_{ij}\left(d_j - b_j\right)}{(\Delta \tmu)_i\,(\Delta\left[\uz\right])_i \,\epsilon_i \,\Phi \,N}\,\,\,, \end{equation}
\par\end{center} 

\noindent where $d_j$ refers to data, $b_j$ the background, $U_{ij}$ is an unfolding matrix connecting the reconstructed variable index $j$ to the true index $i$~\cite{dago}, $\epsilon_i$ is the detection efficiency, $\Delta \tmu$ and $\Delta \left[\uz\right]$ the respective bin widths, $\Phi$ the integrated $\numub$ exposure, and $N$ the number of proton targets in the volume studied.  A particular strength of this cross-section selection and configuration is the unfolding matrix $U$ is entirely independent of assumptions regarding the underlying interaction, and mostly corrects for the detector's measured response to muon kinematics.  At 800 MeV, the energy (angle) resolution is 3.4\% (1.0 deg). Meanwhile, the two-dimensional cross section fully exploits the considerable statistics of the collected sample.  Figure~\ref{fig:dblDiffl} compares the results for $\numub$ CCQE on carbon to a few predictions~\cite{nieves,meucci,amaro,martini}.

The cross section as a function of $\numub$ energy and other differential cross sections are provided in Ref.~\cite{nubDblDiffl}.  Other extracted cross sections, including as a function of incident antineutrino energy are provided in Ref.~\cite{nubDblDiffl}.  This and some other quantities are necessarily model dependent since they can only be calculated assuming information about the underlying interaction. This is the main reason the double differential cross section $\dbldiffl$ is the main result of this work.


\section{$\numub$ NCE cross section}

NCE events are identified in MiniBooNE through observations of low-energy scintillation light.  The prompt Cherenkov signature of muons, electrons and pions are readily rejected from the analysis sample. However, beam-related, low-energy neutral particles produced in neutrino interactions external to the detector (so-called ``dirt" events) may not trigger the veto system and can be accepted by this selection.  This background is roughly the same size as the $\numu$ contribution, at nearly 20\% of the analysis sample.  

\begin{figure}
\includegraphics[scale=0.25]{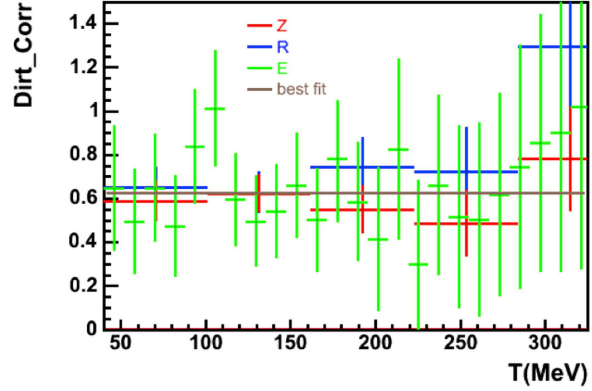} 
\caption{\label{fig:dirtMeas} Summary of measurements of background events produced external to the detector as a function of reconstructed nucleon kinetic energy for the radius (R), beam direction (Z) and energy (E) distributions.  The measurements are shown relative to the nominal and largely uncertain prediction.}
\end{figure}

As in the $\numu$ background to the antineutrino-mode samples, this background can be directly constrained by {\it in situ} measurements.  Energetic and spatial correlations of these events allow their contribution to be checked in a number of distributions: dirt events tend to have a reconstructed vertex at relatively high radius and also in the upstream half of the detector.  They also are correlated with low-energy deposits, and so fitting for the number of dirt events as a function of the reconstructed radius and the vertex position along the beam direction as a function of the observed energy, as well as the energy distribution itself, allows for complementary determinations of this important background.  Figure~\ref{fig:dirtMeas} shows that the results from these analyses are consistent in their indication of a uniform reduction relative to the unconstrained prediction for the dirt contribution.

\begin{figure}
\includegraphics[scale=0.33]{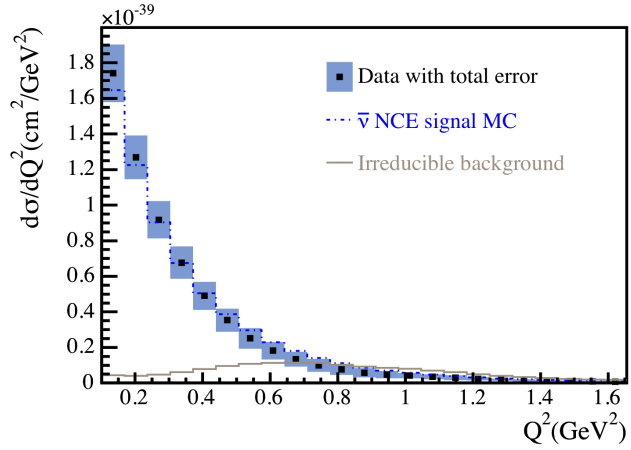} 
\caption{\label{fig:nceXsec} The MiniBooNE $\numub$ NCE flux-averaged differential cross section on mineral oil as a function of momentum transfer compared to a prediction from the RFG. Also shown is the ``irreducible back-ground", which is mostly due to NC$\pi$ events ($\numub + N \to \numub + N' + \pi$), where, due primarily to final-state interactions, the $\pi$ is not observed.}
\end{figure}

A dedicated reconstruction algorithm for NCE events is used to measure the total kinetic energy of all final-state nucleons.  This information can be used to reconstruct the momentum transfer of the interaction: $Q^2 = 2m_N \sum T_{N}$, where $m_N$ is the nucleon mass and $T_N$ its kinetic energy.  Using a calculation similar to Eq.~\ref{eqn:dblDiffl}, the differential cross section with respect to the momentum transfer is presented in Figure~\ref{fig:nceXsec}. 

\section{Combined $\numu$ and $\numub$ cross-section measurements}

\begin{figure}
\includegraphics[scale=0.50]{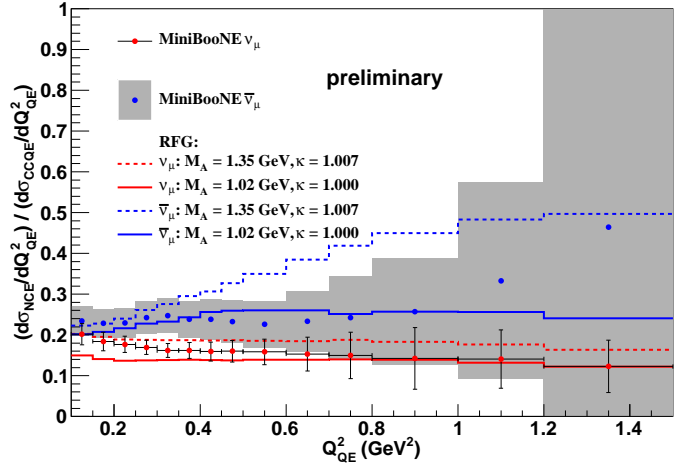} 
\caption{\label{fig:combXsecs} Combined measurements of the NCE and CCQE processes by neutrino species. MiniBooNE data is compared to a few predictions from the RFG, as labeled.  By measuring the ratio of two processes from the same neutrino exposure, common dependences on the $\numu$ and $\numub$ flux vanish.}
\end{figure}

With the $\numub$ measurements of the CCQE and NCE processes described here, an opportunity exists to exploit correlated systematic uncertainties between these data and the analogous $\numu$ results~\cite{qePRD,nuNCE}.  Many opportunities with these four data sets are available, where combined measurements of the same process (CCQE or NCE) across the $\numu$ and $\numub$ results exploit correlated detector uncertainties and combined measurements of the two processes with the same neutrino type takes advantage of common flux uncertainties.  An example of the latter is shown in Figure~\ref{fig:combXsecs}, where ratios of  the momentum transfer distributions for the $\numu$ and $\numub$ NCE and CCQE data are compared to various RFG model predictions.  Note the momentum transfer calculations of CCQE and NCE both involve the assumption of a quasi-elastic process, while the interaction is accessed experimentally by purely leptonic (hadronic) observations for the CCQE (NCE) interaction.  Given the known limitations of the RFG, interpretations of these results must be made with care.  

\section{Conclusion}

The first measurements of $\numub$ CCQE and NCE processes observing a flux of $\numub$ below 1 GeV are presented in this document.  These results, as well as combined measurements with the previously-measured $\numu$ cross sections, will help constrain these processes in current and future searches for neutrino oscillations.  




\bibliographystyle{aipproc}   



\end{document}